\documentclass[conference]{IEEEtran}
\IEEEoverridecommandlockouts
\usepackage{cite}
\usepackage{tikz}
\DeclareUnicodeCharacter{2212}{-}
\usepackage{amsmath,amssymb,amsfonts}
\usepackage{algorithm}
\usepackage{caption}
\usepackage{xcolor}

\usepackage{algpseudocode}
\newcommand{\remove}[1]{}
\usepackage{graphicx}
\usepackage{textcomp}
\usepackage{hyperref}
\usepackage{xcolor}
\usetikzlibrary{positioning, fit}
\usetikzlibrary{fit,arrows,calc,positioning}
\usetikzlibrary{shapes.geometric, arrows}
    \tikzstyle{startstop} = [rectangle, rounded corners, 
minimum width=3cm, 
minimum height=1cm,
text centered, 
draw=black, 
fill=red!30]

\tikzstyle{io} = [trapezium, 
trapezium stretches=true, 
trapezium left angle=70, 
trapezium right angle=110, 
minimum width=3cm, 
minimum height=1cm, text centered, 
draw=black, fill=blue!30]

\tikzstyle{process} = [rectangle, 
minimum width=3cm, 
minimum height=1cm, 
text centered, 
text width=3cm, 
draw=black, 
fill=orange!30]

\tikzstyle{decision} = [diamond, 
minimum width=3cm, 
minimum height=1cm, 
text centered, 
draw=black, 
fill=green!30]
\tikzstyle{arrow} = [thick,->,>=stealth]
\def\BibTeX{{\rm B\kern-.05em{\sc i\kern-.025em b}\kern-.08em
    T\kern-.1667em\lower.7ex\hbox{E}\kern-.125emX}}
\begin{document}

\title{Empowering SMPC: Bridging the Gap Between Scalability, Memory Efficiency and Privacy in Neural Network Inference\\
}
\author{\IEEEauthorblockN{ Ramya Burra}
\IEEEauthorblockA{\textit{IUDX Program Unit, SID, IISc} \\
ramya.burra@gmail.com}
\and
\IEEEauthorblockN{ Anshoo Tandon}
\IEEEauthorblockA{\textit{IUDX Program Unit, SID, IISc} \\
anshoo.tandon@gmail.com}
\and
\IEEEauthorblockN{ Srishti Mittal}
\IEEEauthorblockA{\textit{IUDX Program Unit, SID, IISc} \\
srishti.mittal@datakaveri.org}
}

\maketitle

\begin{abstract}
This paper aims to develop an efficient open-source Secure Multi-Party Computation (SMPC) repository, 
that addresses the issue of practical and scalable implementation of SMPC protocol on machines with moderate computational resources, while aiming to reduce the execution time. We implement the ABY2.0 protocol for SMPC, providing developers with effective tools for building applications on the ABY 2.0 protocol. This article addresses the limitations of the C++ based MOTION2NX framework for secure neural network inference, including memory constraints and operation compatibility issues. Our enhancements include optimizing the memory usage, reducing execution time using a third-party Helper node, and enhancing efficiency while still preserving data privacy. These optimizations enable MNIST dataset inference in just $32$ seconds with only $0.2$ GB of RAM for a 5-layer neural network. In contrast, the previous baseline implementation required $8.03$ GB of RAM and $200$ seconds of execution time. 
\end{abstract}

\begin{IEEEkeywords}
SMPC, ABY2.0, MOTION2NX
\end{IEEEkeywords}

\section{Introduction}

    In today's interconnected and data-driven world, the ability to perform computations while preserving privacy and security is paramount. Privacy is considered a fundamental human right as it balances the need for transparency and accountability with the protection of individual rights~\cite{PrivFundamental}. As technology advances and the digital age evolves, preserving privacy remains a pressing concern that requires ongoing attention and protection. Secure Multi Party Computation~(SMPC) serves as a fundamental tool to address these concerns and facilitates secure data sharing and decision-making across various domains. When implementing SMPC for real-life data, it is essential to consider factors such as the nature of the data, the privacy requirements, the computational resources available, and the specific  tasks to be performed. In this paper, our objective is to tackle the challenge of implementing the SMPC protocol in real world scenarios and at scale on machines with modest computational resources, all while striving to minimize the execution time. Our main goal is to optimize the code, making it more efficient and accessible to a broader audience, ultimately empowering SMPC. We offer neural network inference solutions with semi-honest security~\cite{semi-honest} executed on virtual machines with less than 1 GB RAM. Previous baseline implementation required 8.03 GB RAM on virtual machines for executing the same neural network model. In practical real-world setting, this reduction in RAM usage results in significant cost reduction (in dollar terms)~\cite{amazon-pricing}.

To accomplish this, we modified the C++ based MOTION2NX framework~\cite{braun2021a} and included additional functionality to provide a \emph{resource-optimized} implementation for secure inferencing tasks. We begin by examining MOTION2NX's limitations, such as memory issues and the lack of interoperability between tensor and non-tensor operations. The proposed enhancements include leveraging efficient tensor operations, overcoming the absence of an argmax function using novel approaches, optimizing memory usage, and reducing execution time with the introduction of a third-party Helper node. These improvements aim to enhance the framework's capabilities and efficiency while maintaining data privacy and integrity (refer to Section~\ref{sec:opt-motion2nx} for details). 

In our optimized implementation, it is important to highlight that the memory usage of a standard $N$-layer neural network is determined by its largest layer, i.e., the layer with most parameters. This feature makes our implementation highly scalable as the memory footprint does not grow with the number of layers of the neural network. Moreover, we present an approach to further decrease the memory footprint by splitting the computations for the largest layer without compromising on privacy or accuracy.

We use the data provider framework of SMPC. We consider there are two compute servers (for executing the ABY2.0 SMPC protocol~\cite{ABY2.0}) and two data providers (that actually possess input data and neural network model, respectively). Data providers provide shares to compute servers for computation in this framework (see Section~\ref{dataprovider model} for details). We assume that the readers are familiar with ABY2.0 protocol~\cite{ABY2.0}.

\subsection{Related Work}
Over the past several years, there has been an increased focus on practical application of SMPC to real-world problems.
Here, we elucidate two real-life applications of SMPC.

\paragraph{Secure Auction}In Denmark, farmers sell sugar beets to Danisco. The Market Clearing Prices, which represent the price per unit of the commodity that balances total supply and demand in the auction, play a pivotal role in the allocation of contracts among farmers, ensuring a fair and efficient distribution of production rights. To preserve bid privacy in this process, a three-party SMPC system involving representatives from Danisco, and two other organizations was employed~\cite{smpc-live}. 

\paragraph{Secure Gender Wage Gap Study} Here, a specialized software facilitates data analysis of collaborative compensation for organizations, like the Boston Women's Workforce Council (BWWC) study in Greater Boston~\cite{gender-wage}. This application seamlessly integrates SMPC techniques to ensure collective computation of aggregate compensation data while preserving individual privacy. This approach empowers organizations to collaborate effectively while upholding data privacy. 

We remark that the above two described SMPC applications are not memory or computationally intensive. On the other hand, in this paper, we present the modifications and functional additions to the MOTION2NX framework for practical implementation of secure neural network inference task that is \emph{both} memory intensive and computationally intensive. These modifications and updates are a step towards secure disease prediction (see~\cite{secure-nn} for secure medical image analysis) where one party provides secret shares of medical images while the other party provides secret shares of a pre-trained neural network model. 

\subsection{Our Contributions}
The following is the list of our contribution, specifically to MOTION2NX setup.
\begin{itemize}
\item Extending MOTION2NX to the setup where data providers and compute servers are separate entities.
\item Creating an argmax function for multiple ($ >2$) inputs.
\item Enabling writing of output shares to files and preventing reconstruction of the output at the compute servers. The compute servers send their respective final shares to the output owners.
\item Inter-operability of tensor and non-tensor operations.
\item A new optimized Helper node algorithm, working in conjunction with the ABY2.0 protocol, for efficient matrix multiplication.
\item Our optimized 5-layer neural network inference requires about 0.2 GB of RAM, where as the original non-optimized version requires $8.03$ GB of RAM. This implies over $40 \times$ reduction in RAM usage. 
\item For obtaining realistic numbers, we opted to deploy the
Compute servers and the Helper node on the cloud. Set-
ting up and operating these systems require considerable amount of time due to installation of necessary dependencies and compilation of binaries in each of the machines. To simplify this process, we offer the
docker images which contains all the essential dependencies and compiled binaries of the new enhancements we contributed.
The SMPC Compute servers can now be easily brought up as docker containers from these docker images 
on any machine with docker installation.

{\color{black}The source code of our optimized implementations along with docker images is available at~\url{https://github.com/datakaveri/iudx-MOTION2NX}. }
\end{itemize}
\section{Preliminaries}
In this section, we provide preliminary details of our secure neural network inferencing implementation.
\subsection{Framework for Implementation}
We consider MOTION2NX, a C++ framework for generic mixed-protocol secure two-party computation in our paper. The following are the features of baseline MOTION2NX.
\begin{itemize}
     \item Assumes data providers are a part of compute servers
    \item No intermediate values are reconstructed
    \item Assumes either of the compute servers as output owners
    \item Output is reconstructed in clear and shared with the output owner 
\end{itemize}
\paragraph*{Tensor and non-Tensor variants}
\label{tensor,non-tensor}
MOTION2NX offers non-optimized secure functions that compile the descriptions of low-level circuits, which are referred to as non-tensor operations. MOTION2NX also provides optimized building blocks that directly implement common high-level operations, which are referred to as tensor operations. The tensor operations are more computationally efficient than the primitive operations. They use a specialized executor to evaluate the tensor operations sequentially while parallelizing the operations themselves with multi-threading and SIMD operations. 

We discuss the details of our proposed enhancements and optimizations in Section~\ref{sec:opt-motion2nx}.

\subsection{$N$-layer Neural Network}

Our optimizations in MOTION2NX enable us to implement deep neural networks with relatively large number of layers. For illustrative purposes, we only present the details for 2-layer and 5-layer neural networks on MNIST dataset. A similar procedure can be adapted for any other dataset on their pretrained models with multiple neural net layers.

The input to the neural network is a real-valued vector of size $784 \times 1$. The output is a boolean vector of size $10$, where only one element is set to $1$. Each element in the output vector corresponds to an index from $\{0, 1, 2, \ldots, 9\}$. If the $j^{th}$ element in the vector is $1$, it indicates that the predicted label is $j$. The dimensions of weights and biases used in our implementation are listed below.

\subsubsection{2 layer neural network dimensions}
\label{2layer-dimensions}
\begin{itemize}
\item Layer 1 : Weights $(256 \times 784)$, bias $(256 \times 1)$ 
\item Layer 2 : Weights $(10 \times 256 )$, bias $(10 \times 1)$
\end{itemize}
\subsubsection{5 layer neural network dimensions}
\label{5layer-dimensions}
\begin{itemize}
\item Layer 1 : Weights $(512 \times 784)$, bias $(512 \times 1)$
\item Layer 2 : Weights $(256 \times 512)$, bias $(256 \times 1)$
\item Layer 3 : Weights $(128 \times 256)$, bias $(128 \times 1)$
\item Layer 4 : Weights $(64 \times 128)$, bias $(64 \times 1)$
\item Layer 5 : Weights $(10 \times 64)$, bias $(10 \times 1)$
\end{itemize}

Algorithm~\ref{alg:neural_net_inferencing} describes a simple two-layer secure neural network inference implementation with ReLU activation. This algorithm takes ABY2.0 shares of input data (image), neural network weights, and biases as inputs and produces ABY2.0 shares of the predicted label as output.

We recall that ABY2.0 shares consist of a pair comprising a public share and a private share~\cite{ABY2.0}. For instance, the ABY2.0 shares of an input variable $y$ associated with server $i$ are represented as a pair consisting of $\Delta_y$ and $[\delta_{y}]_i$. Note that $\Delta$ and $\delta$ represent public share and private share respectively. 
\begin{algorithm}
    \caption{Neural Net inferencing task with ReLU activation function at compute server-$i$, $i \in \{0,1\}$}
    \label{alg:neural_net_inferencing}
    \begin{algorithmic}[1]
        \Require Input image shares $x^i$, weight shares $w^i_1,w^i_2$, bias shares $b^i_1,b^i_2$. All the above shares are vectors in the form of ABY2.0 shares
        \Ensure Shares of predicted class label $\hat{y}^i$
        \State Compute first layer: \\
        \quad $z^i_1 = $SecureAdd(SecureMul$(w^i_1,x^i) ,b^i_1)$ \\
        \quad $h^i_1 = $SecureReLU$(0, z^i_1)$ 
        \State Compute Second layer: \\
         \quad $z^i_2 = $SecureAdd(SecureMul$(w^i_2,h^i_1) ,b^i_2)$ 
        \State Compute predicted class label: \\
        \quad $\hat{y}^i= $SecureArgmax($ z_2$) 
        \State \textbf{Return} $\hat{y}^i$

    \end{algorithmic}
\end{algorithm}
The secure functions listed in Algorithm~\ref{alg:neural_net_inferencing} were provided by MOTION2NX framework. We used ABY2.0 arithmetic protocol for multiplication and addition. We used Yao protocol to perform the ReLU function, as Yao performs better for comparison operations. Before performing Yao operations we convert ABY2.0 arithmetic shares to Yao shares with the help of MOTION2NX functions. Post the ReLU function we convert the Yao shares to ABY2.0 arithmetic shares with the help of MOTION2NX inbuilt functions. We choose boolean ABY2.0 protocol to implement the argmax function.

Algorithm~\ref{alg:neural_net_inferencing}, unfortunately, couldn't be executed completely in MOTION2NX using its built-in functions.\remove{ We now briefly delve into the encountered challenges. 
It's important to note that MOTION2NX offers two distinct sets of functions: one for non-tensor operations, which represents an unoptimized version, and the other for tensor operations that are optimized for machine learning tasks.} In a single MOTION2NX instance, only one of the tensor variant or non-tensor variant could be utilized. Regrettably, the tensor variant lacked an argmax function, which prevented us from implementing Algorithm~\ref{alg:neural_net_inferencing} using its standard functions. Nevertheless, we were able to address this issue by applying the optimizations outlined in Section~\ref{Optimizations:Argmax,output_shares}, which provides further insights into this matter.

We remark that the steps outlined in Algorithm~\ref{alg:neural_net_inferencing} can be readily extended to provide secure inferencing results for a general fully-connected neural network with $N \left(> 2\right)$ layers.
\vspace{0.1in}

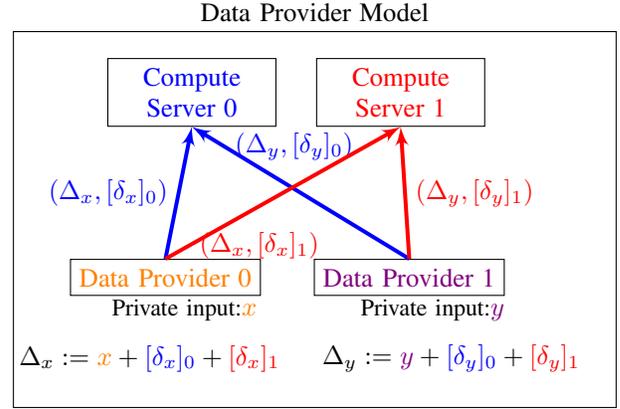
\tikzstyle{space} = [rectangle]
\tikzstyle{p0} = [rectangle, draw]
\tikzstyle{p1} = [rectangle, draw]
\tikzstyle{dp0} = [rectangle, draw]
\tikzstyle{dp1} = [rectangle, draw]
\tikzstyle{l} = [draw, -latex',thick]
\begin{tikzpicture}[auto]
    \node[space](S) {}; 
    \node [p0, right=of S, xshift=-10pt, text width=2cm, align=center] (P0) {\textcolor{blue}{Compute Server 0}};
    \node [p1, right=15pt of P0, text width=2cm, align=center] (P1) {\textcolor{red}{Compute Server 1}}; 
        \node [dp0, below=50pt of P0, xshift=-10pt] (dP0) {\textcolor{orange}{Data Provider 0}};
    \node [dp1, right=20pt of dP0] (dP1) {\textcolor{violet}{Data Provider 1}}; 
    \node (tx) at (1.8,-2.9) {{\small Private input:}{\textcolor{orange}{{$x$}}}};
    \node at (5.1,-2.9) {{\small Private input:}{\textcolor{violet}{{$y$}}}};
        \draw[<-, >=latex', line width=1.5pt, blue] (P0.south) -- node[pos=0.15, right] {$(\Delta_y,\textcolor{blue}{[\delta_y]_0})$}(dP1.north);
         \draw[<-, >=latex', line width=1.5pt, red] (P1.south) -- node[right] {$(\Delta_y,\textcolor{red}{[\delta_y]_1})$}(dP1.north);
          \draw[<-, >=latex', line width=1.5pt, blue] (P0.south) -- node[left] {$(\Delta_x,\textcolor{blue}{[\delta_x]_0})$}(dP0.north);
         \draw[<-, >=latex', line width=1.5pt, red] (P1.south) -- node[pos=0.9, right]{$(\Delta_x,\textcolor{red}{[\delta_x]_1})$}(dP0.north);
            \node[below=15pt of dP0, xshift=-6pt] (s1){$\Delta_x:=\textcolor{orange}{x}+\textcolor{blue}{[\delta_x]_0}+\textcolor{red}{[\delta_x]_1}$};
    \node[right=9pt of s1] (s2){$ \Delta_y:=\textcolor{violet}{y}+\textcolor{blue}{[\delta_y]_0}+\textcolor{red}{[\delta_y]_1}$};
    \node[fit=(S)(P0)(P1)(dP0)(dP1)(tx)(s2), draw, inner sep=10pt, rectangle, label=Data Provider Model] (captionbox) {\label{dataprovider-model}};
\end{tikzpicture}

\subsection{Reconstructing the Predicted label in clear}
Algorithm~\ref{alg:reconstruction} outlines the process of deriving the predicted label from boolean output shares. With MNIST, the label is a number from $0$ to $9$. The output here is a boolean vector of 10 elements indexed from $0$ to $9$, with only one element set to 1; a $1$ in the $j^{th}$ element signifies the predicted label as $j$.

\begin{algorithm}
    \caption{Reconstructing the Predicted label in clear}
    \label{alg:reconstruction}
    \begin{algorithmic}[1]
        \Require Boolean ABY2.0 shares from both the servers $\hat{y}^0,\hat{y}^1$
        \Ensure Predicted class label $Y$
        \For{$\text{j} = 0$ to $9$} 
         \State $\hat{y}[j]=\Delta_{\hat{y}[j]} \oplus [\delta_{\hat{y}[j]}]_0 \oplus [\delta_{\hat{y}[j]}]_1$ 
          
        \EndFor
        \State Initialize index $k \gets 0$
        
        \While{$k < 10$ and $\hat{y}[k] = 0$}
            \State Increment index $k \gets k + 1$
        \EndWhile
        
        \State \textbf{Return} $k$ \Comment{The predicted label}

    \end{algorithmic}
\end{algorithm}

\subsection{Data Provider Model}
\label{dataprovider model}
In this model the data providers~(Image provider and Model provider) create shares of their private data and communicate them with the compute servers for further computation. Compute servers perform the inferencing task and send the output shares to the Image provider. Compute servers are unaware of the clear output result. 
For secure inferencing task, we consider that the neural network model is pretrained and is proprietary to a model provider. Similarly, the image for the inferencing task is private to image data provider. 


\section{Enhancements, Optimizations, and Feature Additions to MOTION2NX}
\label{sec:opt-motion2nx}
Before we proceed with a discussion of the enhancements proposed in our paper, let's first examine the limitations of MOTION2NX.

\paragraph*{Limitations}
\begin{itemize}
    \item No provision to implement data provider model
    \item Output is always reconstructed at the compute server(s)
    \item Memory issues : A simple 2-layer neural network \\
    inference requires about 3.2 GB of RAM
    \item No interoperatability between tensor and non-tensor operations.
\end{itemize}

\subsection{Floating point to fixed point and back, and potential errors introduced}
\label{encode,decode,fractional-bits}
Motion2NX internally works in the uint operations. We had to design appropriate encode functions to convert real numbers to uint depending on the number of fractional bits we wish to consider. Also, during reconstruction the uint numbers had to be appropriately mapped back to real numbers. This process introduces an error due to fixed point arithmetic. Interestingly, for inferencing task on the MNIST data set we see that for fractional bits as small as 6, no considerable error is introduced. Specifically, the accuracy for secure computation with 6 fractional bits matches the accuracy of the inferencing task in the floating point world when tested with 1000 test images in the MNIST dataset. To understand the error introduced due to secure computation and uint operations, we compared the cross entropy loss of the output vector at layer 2 for a 2 layer neural network, before the argmax operation. We see that the cross entropy loss decreases as we increase the fractional bits upto 24. We also observed that a further increase in fractional bits increases the cross entropy loss. Also, beyond the threshold accuracy of the SMPC model decreases drastically. This inference is in tune with~\cite[paragrah 4, Section 5.1.1]{ABY3} that talks about the adverse impact of increase in fractional bits due to wrap around error during the truncation operation. 
\subsection{Implementing Argmax and writing output shares}
\label{Optimizations:Argmax,output_shares}
MOTION2NX can implement the functions (matrix multiplication, add etc.) using tensor and non-tensor operations. 
As discussed in Section~\ref{tensor,non-tensor}, tensor operations are optimized versions that enable faster calculations. Therefore, to perform an inference operation on a neural net, MOTION2NX authors recommend using tensor operations. Also, practically we verified that inferencing task on MNIST dataset takes around 1 minute using tensor operations. A simple multiplication operation of two numbers in a non-tensor version takes about 1.5sec. An inferencing task on MNIST data set on a two layer neural network has operations of the order $\approx 10^6$. Clearly, the tensor operations are much more efficient compared to the non-tensor operations.

To implement the neural net we need to have an argmax function (providing the index of the maximum value in a vector) for appropriate classification of the MNIST image. However, MOTION2NX has no provision to implement the argmax function using the tensor operations. Although there are functions to compute max and argmax using non-tensor operations for two inputs, these functions cannot be used in the inferencing task as \emph{there is no provision to interlink tensor and non-tensor operations in MOTION2NX}. We modified MOTION2NX to store the output shares of the neural net just before executing the argmax function, while ensuring that the corresponding clear values of the output shares are never reconstructed in the code. This feature is an improvement over the existing framework as it enables us to use tensor and non-tensor operations together in a sequential manner. The output shares of the last layer of the neural net are fed to the non-tensor implemetation of the argmax function to obtain the required shares of the predicted label. This argmax function operates on a vector with multiple $\left( >2 \right)$ inputs, and internally uses the in-built max and argmax functions for two inputs in a recursive fashion for generating the output shares of the predicted label. 

In our implementation, we pinpointed the stage at which output shares become accessible, just before the reconstruction step. Note that in the original MOTION2NX framework, the compute servers initiate the transmission of their private share for the purpose of reconstruction.
We modified the MOTION2NX framework to halt the broadcast of the private shares, and instead save the local public and private shares to respective files at the two compute servers. It is important to note that the private shares of each server remain local (not shared with the other server), ensuring no loss of privacy.

\subsection{Optimizing Memory Requirement}
In the context of secure inferencing tasks using the MNIST dataset, we observed a substantial RAM requirement of approximately 3.2 GB (per server instance) for a two-layer neural network. In practical terms, this RAM demand poses a significant obstacle to performing inferencing tasks on a resource constrained machine. This issue is further aggravated when we work with a neural network with relatively large number of layers. To address this challenge, our primary objective was to reduce the memory requirement, thereby facilitating the use of more complex neural networks.

Notably, for a multi-layer neural network, MOTION2NX constructs the entire end-to-end circuit in a single step and retains the RAM memory until the entire execution has been completed. Our approach, as outlined in Section~\ref{Optimizations:Argmax,output_shares}, involves writing the shares of intermediate results in respective files while ensuring that no reconstruction is performed. This significantly reduces the RAM requirement for multi-layer neural network inferencing tasks. 

In the specific context of a 2-layer neural network inferencing model~(with dimensions detailed in Section~\ref{2layer-dimensions}), the highest memory requirement arises from the matrix multiplication in layer-1. To mitigate this, we perform intra-layer optimization where the matrix multiplication task is implemented in smaller segments (or ``splits"), resulting in a proportional decrease in the average RAM requirement. It is evident from Table~\ref{Table:Two_Layer_NN} that the average RAM requirement scales down almost linearly with the number of splits employed.  

 \begin{table} 
 \centering
      \begin{tabular}{|c|c|c|} 
        \hline
       {Splits} & {RAM requirement} & {Execution time}\\
      
\hline
  No intra-layer split & 3.2 GB & 34 seconds \\
\hline
 Layer 1: 2 splits & 1.6419 GB & 38 seconds \\
\hline
 Layer 1: 4 splits & 0.888 GB & 43 seconds \\
\hline
 Layer 1: 8 splits & 0.4527 GB & 47 seconds \\
\hline
 Layer 1: 16 splits & 0.253 GB & 50 seconds \\
\hline
 Layer 1: 64 splits & 0.09888 GB & 73 seconds \\
\hline
  \end{tabular}

\caption{Compute servers run on the same machine for 2 layer NN. Note that there is no intra-layer split for layer2 in the above.}\label{Table:Two_Layer_NN}
\vspace{-0.1in}
\end{table}

\subsection{Optimizing Execution Time using Helper Node Algorithm} 
\label{opt-exec-time}
After successfully reducing the average RAM requirement, our next objective was to optimize the execution time. 
Here, we implemented compute servers on different machines but on same LAN for realistic execution times. The numbers are detailed in Tables~\ref{table:2layer-split} and~\ref{table:5layer-split}. For secure matrix multiplication, the execution time is significantly impacted by the use of oblivious transfers (OTs) which occur behind the scenes. To address this issue, we introduce a semi-honest third-party Helper node that eliminates the need for OTs during matrix multiplication. 

As a result of implementing the Helper-node algorithm, the execution time for the inference operation is reduced from 77 sec (for the baseline implementation with no intra-layer optimization on a 2-layer nural network) to 11 sec. 
Additionally, the RAM requirement when utilizing the Helper node algorithm is only 0.134 GB~(see Table~\ref{table:2layer-split}).

Note that we refer to this algorithm as ``ABY2.0 with a Helper node", drawing inspiration from the implementation of Beaver triples produced by a third-party helper in~\cite{helper-inspiration}. 
The primary objective of our Helper node is to eliminate the need for Oblivious Transfers (OTs) in both the online and setup phases of ABY2.0. 
In this context, we specifically delve into how the Helper node's algorithm modifies the ABY2.0 Multiplication Protocol, denoted as Protocol MULT$\left(⟨a⟩, ⟨b⟩\right)$ in~\cite[Section 3.1.3]{ABY2.0}.

In summary, Protocol MULT$\left(⟨a⟩, ⟨b⟩\right)$ takes ABY2.0 shares of $a$ and $b$ from two parties as inputs. It performs operations to generate output shares representing the product $a \times b$. Importantly, in this protocol, neither party possesses knowledge of the clear output unless they engage in communication to share their respective output secret shares.

It's worth noting that Protocol MULT $(⟨a⟩, ⟨b⟩)$ relies on an OT-based setupMULT during the pre-processing phase. Our specific goal is to eliminate this setupMULT and replace it with a more efficient approach using helperNODE. Here, the objective is to calculate \begin{equation} \label{eqn:helper}
\delta_{ab}=[([\delta_a ]_0 +[\delta_a ]_1 )([\delta_b ]_0+ [\delta_b ]_1 )],
\end{equation} and share the additive components of this expression with both server nodes. In this process, party 0 conveys $[\delta_a]_0$ and $[\delta_b]_0$ to the Helper node through a reliable channel, while party 1 conveys $[\delta_a]_1$ and $[\delta_b]_1$ to the Helper node via another reliable channel. Subsequently, the Helper node computes $\delta_{ab}$ and distributes additive shares to both parties. It is crucial to note that the Helper node is \textit{incapable of reconstructing the actual values of $a$ and $b$ as it possesses no knowledge of $\Delta_a$ and $\Delta_b$}. Additionally, we assume that both the compute server parties and the Helper node operate in a semi honest fashion. 
\remove{
adhere to the following properties.
\begin{itemize}
\item  Code integrity
\item  Data integrity and
\item  Data confidentiality
\end{itemize}
The aforementioned characteristics guarantee the confidentiality of data from both the servers and the Helper node, preventing any unintended disclosures. }It is important to emphasize that if the information from the Helper node is shared with either of the two servers, it jeopardizes the data privacy of the input data. To uphold these properties, one may consider running the compute servers and the Helper node on separate secure enclave machines (see~\cite{secure-enclave} for details on the secure enclave machines).

In Algorithm~\ref{alg:mult_hepler}, we provide an overview of the multiplication algorithm using the Helper node. This algorithm needs the procedure HelperNODE in the setup phase. \remove{Specifically, we propose a new method to implement setupMULT function from ABY2.0 paper to save on the average memory requirement and execution time of the inferencing task. }To enhance readability, Algorithm~\ref{alg:mult_hepler} addresses the efficient implementation of two scalar values, $a$ and $b$. However, we carry forward this very idea into our optimized implementation for matrix multiplication.

\begin{algorithm}
   \caption{Protocol HELPERMULT$(⟨a⟩, ⟨b⟩)$}
   \label{alg:mult_hepler}
   \begin{algorithmic}[1]
      \State \textbf{Setup Phase:} 
      \State $\text{P}_i$ for $i \in  {0, 1}$ samples random $[\delta_y ]_i \in_{R} \mathbb{Z}_2^{64} $.
      \State Parties execute \Call{helperNODE}{$[\delta_a ] , [\delta_b ]$} to obtain $[\delta_{ab}]_i$.
      \State \textbf{Online Phase:} 
      \State  $\text{P}_i$ for $i \in  {0, 1}$ locally computes $[\Delta_y ]i = i  \Delta_{a}\Delta_{b} − \Delta_a [\delta_b ]_i − \Delta_b [\delta_a ]_i + [\delta_{ab} ]_i + [\delta_y ]_i$ and sends to $\text{P}_{1-i}$.
      \State $\text{P}_i$ for $i \in  {0, 1}$ computes $\Delta_y = [\Delta_y ]_0 + [\Delta_y ]_1$ . 
   \end{algorithmic}
\begin{algorithmic}[1]
   \Procedure{helperNODE}{$[\delta_a ] , [\delta_b ]$}
      \State $\text{P}_i$ for $i \in  {0, 1}$ send $([\delta_a ]_i, [\delta_b ]_i)$ to the Helper node.
      \State Helper node computes equation~(\ref{eqn:helper}). 
      \State Helper node creates arithmetic shares $([\delta_{ab}]_0,[\delta_{ab}]_1)$, such that $\delta_{ab}:=[\delta_{ab}]_0+[\delta_{ab}]_1$
      \State Helper node sends $[\delta_{ab}]_i$ to $\text{P}_i$ for $i \in  {0, 1}$
   \EndProcedure
   \end{algorithmic}
\end{algorithm}

\section{Numerical Evaluation}
\label{Numerical-evaluation}
 \begin{table}
 \vspace{-0.1in}
  \centering
      \begin{tabular}{|c|c|c|}
        \hline
       {Splits} & {RAM requirement} & {Execution time}\\
      
\hline
No intra-layer split & 3.2 GB & 77 seconds \\
\hline
 Layer 1: 8 splits & 0.4527 GB & 100.3 seconds \\
\hline
 Layer 1: 16 splits & 0.253 GB & 108.45 seconds \\
\hline
 Layer 1: 64 splits & 0.09888 GB & 121.53 seconds \\
\hline
\hline
\textbf{Helper node} & 0.134GB & 11 seconds \\
 \hline
  \end{tabular}
\caption{Compute servers and Helper node are run on different machine on same LAN for 2-layer neural net. Note that there is no intra-layer split for layer2 in the above.}
\label{table:2layer-split}
 \centering
      \begin{tabular}{|c|c|c|}
        \hline
       {Which layers are split, } & {RAM requirement} & {Time of execution}\\
       number of splits & & \\
  \hline
  No intra-layer split & 8.03 GB & 200 seconds \\
  \hline
  Layer 1: 8 splits & 0.894 GB& 212 seconds \\
  Layer 2: 4 splits && \\
  Layer 3: 2 splits && \\
  Layer 4,5: No split & & \\
 \hline
  Layer 1: 16 splits & 0.461 GB & 210 seconds \\
  Layer 2: 8 splits & & \\
  Layer 3: 4 splits & & \\
  Layer 4: 2 splits & & \\
  Layer 5: No split & & \\
\hline
\hline
\textbf{Helper node} & 0.2 GB & 32 seconds \\
\hline
  \end{tabular}
  \caption{Compute servers and Helper node are run on different machine on same LAN for 5-layer neural net.}
  \label{table:5layer-split}

 \centering
      \begin{tabular}{|c|c|}
        \hline
Server 0 &
Azure: b1s 1vcpu, 1 GB RAM, 30 GB SSD \\
\hline
Server 1 & 
AWS: t2.micro 1vcpu, 1 GB RAM, 30 GB SSD \\
\hline
Helper node  & 
AWS: t2.nano 1vcpu, 0.5 GB RAM, 30 GB SSD \\
\hline
Image provider & Personal laptop \\
\hline
Weights provider & Personal laptop \\
 \hline
  \end{tabular}
  \caption{Cloud configuration}
  \label{table:cloud-specs}

 \centering
      \begin{tabular}{|c|c|c|}
        \hline
       {Splits} & {RAM requirement} & {Time of execution}\\
         \hline
    Layer 1: 16 splits & 0.253 GB & 200 seconds \\
        \hline
      Layer 1: 64 splits & 0.0988 GB & 280 seconds \\
   \hline
   \hline
 \textbf{Helper node} & 0.134 GB & 12 seconds \\
 \hline
      \end{tabular}
\caption{Compute servers and Helper node are run on cloud for 2-layer neural net. Note that there is no intra-layer split for layer2 in the above.}
\label{table:2layer-cloud}

 \centering
      \begin{tabular}{|c|c|c|}
        \hline
       {Which layers are split, } & {RAM requirement} & {Time of execution}\\
       number of splits & & \\
 \hline
  Layer 1: 16 splits & 0.461 GB & 525 seconds \\
  Layer 2: 8 splits & & \\
  Layer 3: 4 splits & & \\
  Layer 4: 2 splits & & \\
  Layer 5: No split & & \\
\hline
\hline
\textbf{Helper node} & 0.2 GB & 34 seconds \\
\hline
  \end{tabular}
  \caption{Compute servers and Helper node are run on cloud for 5-layer neural net.}
\label{table:5layer-cloud}
\vspace{-0.1in}
  \end{table} 
Tables~\ref{table:2layer-split} and~\ref{table:5layer-split} outline the execution time for 2-layer and 5-layer neural networks, respectively, when the compute resources are run on the same LAN.  In practice, the compute servers are typically hosted on the cloud (on different LANs). To determine the execution time, we deployed compute server~0 on Microsoft Azure cloud and compute server~1 (and the Helper node) on AWS cloud. The image provider and weights/model provider run on separate local machines. See Table~\ref{table:cloud-specs} for details of the cloud configurations.

The image provider's terminal accepts a handwritten digit and converts it into a $28 \times 28$ .csv file, which is then flattened to a $784 \times 1$ format. The model provider possesses the weights and biases of a pretrained model. Both the image provider and model provider generate ABY2.0 shares of their respective private data and send these shares to the compute servers hosted on the cloud. Following the sharing of these secret shares, the image provider awaits the output shares of the predicted image label. After the execution of the secure inferencing task at the two compute servers, these servers send their respective output shares back to the image provider for the reconstruction of the predicted image label in plain text. Tables~\ref{table:2layer-cloud} and~\ref{table:5layer-cloud} list details of the 2-layer and 5-layer neural networks, respectively, when compute servers are run on cloud. As expected, the execution time with compute servers running on cloud machines (on different LANs) is higher compared to the case where compute servers are part of the same LAN (compare Tables~\ref{table:2layer-split},~\ref{table:5layer-split} with Tables~\ref{table:2layer-cloud},~\ref{table:5layer-cloud}). 

We remark that the accuracy of our secure neural network inferencing implementation (using 64 bit fixed-point arithmetic including 13 bits for representing the fractional part) on the MNIST dataset was roughly similar to the corresponding accuracy obtained using a python floating-point implementation.

\section{Model known to compute servers}

We recognized the possibility of situations where neural network model parameters might not be private but rather considered common knowledge. However, the image provider still values the privacy of its data. In order to address such scenarios, we have introduced a new functionality where common knowledge variables are treated as known unencrypted values. To achieve this, we have introduced two operations, namely ``ConstantMul" and ``ConstantAdd", within the MOTION2NX framework. These operations were not previously available in the framework. We discuss these algorithms in Algorithm~\ref{alg:const_mult} and~\ref{alg:const_add}, respectively.

When dealing with neural networks where weights are considered common knowledge, we observed that the inference time and RAM requirements are significantly reduced when compared with the baseline~(where the model is private to one of the data providers). When the model is known to both the compute servers, for a two-layer neural network, the inference time is approximately 13 seconds with a RAM requirement of 0.134 GB. In the case of a five-layer neural network, the inference time is around 34 seconds, and the RAM requirement is 0.2 GB. These numbers are similar to the corresponding numbers obtained using the helper node. When the model is known to both the compute servers, this reduction in execution time is expected since the overhead of performing OTs during secure multiplication between two private values is eliminated. 

Let $a$ be the private data, $b$ be the known data. In the following we use ABY2.0 shares of $a$ and uint64 equvivalent of $b$ with $f$ fractional bits. Let $y = a \times b$. It's important to highlight that during the online phase, after the multiplication of the shares by the constant value, we transformed the ABY2.0 shares into the GMW arithmetic shares. Note that the GMW shares are represented by $Y_i$ in Algorithm~\ref{alg:const_mult}. The transformation to GMW shares was needed to prevent the occurrence of wrap-around errors during the truncation operation performed on $Y_i$ (see~\cite[paragrah 4, Section 5.1.1]{ABY3} for a discussion of wrap-around errors during the truncation of respective shares). The truncation operation ensures that after the multiplication operation, the fractional bits continue to be represented by $f$ least significant bits.

\begin{algorithm}
   \caption{Protocol ConstantMULT$(⟨a⟩, b)$}
   \label{alg:const_mult}
   \begin{algorithmic}[1]
  \State \textbf{Setup Phase:} 
   $P_i$ for $i \in \{0, 1\}$ samples $[\delta_y]_i \in_{\mathbb{R}} Z_{2^{64}}$
    \State $P_i$ updates $[\delta_a]_i$ with $[\delta_a]_i \times b$ 
\State \textbf{Online Phase:} 
   $P_i$ updates $\Delta_a$ with $\Delta_a \times b$
    
    \State $P_0$ locally generates $Y_0 = 0 \times \Delta_a - [\delta_a]_0$
    \State $P_1$ locally generates $Y_1 = 1 \times \Delta_a - [\delta_a]_1$
    
    \State $P_i$ for $i \in \{0, 1\}$ performs truncation operation: $Y_i = \frac{Y_i}{2^f}$
    
    \State $P_i$ locally computes $[\Delta_y]_i = Y_i + [\delta_y]_i$ and sends to $P_{1-i}$
    
    \State Both $P_0$ and $P_1$ calculate $\Delta_y = [\Delta_y]_0 + [\Delta_y]_1$
   \end{algorithmic}
\end{algorithm}

\begin{algorithm}
   \caption{Protocol ConstantAdd$(⟨a⟩, b)$}
   \label{alg:const_add}
   \begin{algorithmic}[1]
  \State \textbf{Setup Phase:} 
$P_i$ for $i \in \{0, 1\}$ computes $[\delta_y ]_i = [\delta_a ]_i$
\State \textbf{Online Phase:} 
 Both $P_0$ and $P_1$ compute $[\Delta_y ] = [\Delta_a ]+b$
   \end{algorithmic}
\end{algorithm}
\section{Conclusion and Future Work}
We modified and enhanced the MOTION2NX framework to bridge the gap between scalability, memory efficiency and privacy. In particular, we optimized the memory usage, reduced the execution time using a third-party Helper node, and
enhanced the efficiency while still preserving data privacy. These
optimizations enable MNIST dataset inference in just 32 seconds
with only 0.2 GB of RAM for a 5-layer neural network. In
contrast, the previous baseline implementation required 8.03 GB of
RAM and 200 seconds of execution time

After successfully deploying a neural network inference implementation with $N$-layers, we proceeded to implement CNN inference using the optimized strategies outlined in Section~\ref{Optimizations:Argmax,output_shares}. These optimizations yielded remarkable results in terms of reduced inference time and lower average memory consumption. Our next objective is to optimize the secure neural network \emph{training} implementation. 
{\color{black}\section*{Acknowledgements}
\remove{We extend our heartfelt gratitude to the National Science Council (NSC) for their invaluable support throughout this project, made possible through grant number --grant number.} We want to thank all the interns (Udbhav, Rishab, Rashmi, and Shreyas) who extended their support to this project. We also thank Abhilash for helping us setup the cloud instances.}
\bibliographystyle{plain} 
\bibliography{references} 

\end{document}